\documentclass[12pt,a4paper, final]{iopart}

\usepackage{iopams}  
\usepackage{graphicx}
\usepackage[breaklinks=true,colorlinks=true,linkcolor=blue,urlcolor=blue,citecolor=blue]{hyperref}

\begin{document}

\title[Resonance Raman scattering and electron energy loss spectra of   $\rm{MoS_2}$]{Resonance Raman scattering and $ab$ $initio$ calculation of electron energy loss spectra of   $\rm{MoS_2}$ nanosheets }

\author{Anirban Chakraborti}
\address{School of Computational \& Integrative Sciences, Jawaharlal Nehru University, New Delhi-110067, India}
\ead{anirban@jnu.ac.in}

\author{Arun Singh Patel }
\address{School of Computational \& Integrative Sciences, Jawaharlal Nehru University, New Delhi-110067, India}
\ead{arunspatel.jnu@gmail.com}

\author{Pawan K. Kanaujia}
\address{Nanophotonics Laboratory, Department of Physics, Indian Institute of Technology Delhi, New Delhi-110016, India}
\ead{kanaujiapk09@gmail.com}

	\author{Palash Nath} 
	\address{Department of Physics, University of Calcutta, 92 APC Road, Kolkata-700009, India}
	\ead{palashnath20@gmail.com}
	
	\author{G.Vijaya Prakash} 
	\address{Nanophotonics Laboratory, Department of Physics, Indian Institute of Technology Delhi, New Delhi-110016, India}
	\ead{prakash@physics.iitd.ernet.in}
	
	\author{Dirtha Sanyal} 	
	\address{Variable Energy Cyclotron Centre, 1/AF Bidhannagar, Kolkata-700064, India}
	\ead{dirtha@vecc.gov.in}

\begin{abstract}
The presence of electron energy loss (EELS) peak is proposed theoretically in molybdenum disulfide ($\rm{MoS_2}$) nanosheets. Using density functional theory simulations and calculations, one EELS peak is identified in the visible energy range, for $\rm{MoS_2}$ nanosheets with molybdenum vacancy. Experimentally, four different laser sources are used for the Raman scattering study of $\rm{MoS_2}$ nanosheets, which show two distinct Raman peaks, one at 385 cm$^{-1}$ ($E_{2g}^{1}$) and the other at 408 cm$^{-1}$ ($A_{1g}$). In the cases of three laser sources with wavelengths 405 nm (3.06 eV),  632 nm (1.96 eV) and  785 nm (1.58 eV), respectively, the intensity of $E_{2g}^{1}$  Raman peak is more than the $A_{1g}$ Raman peak, while in the case of excitation source of 532 nm (2.33 eV), the intensity profile is reversed and $A_{1g}$ peak is the most intense. Thus a resonance Raman scattering phenomenon is observed  for 532 nm laser source. 

\end{abstract}

\vspace{2pc}
\maketitle

\section{Introduction}

The discovery of graphene in 2004 provided a novel way of studying two dimensional layered structure materials \cite{Kostya2004, Kin2010, Branimir2011, Jakob2012, Nath2014275, Jiadan2013, Shakya2016}. In the recent few years, it has been possible to synthesize different kind of two dimensional nanomaterials of metal chalcogenides, in which d-electrons' interactions can give rise to new physical phenomena. 	Among these materials, the transition-metal dichalcogenide semiconductor $\rm{MoS_2}$ has attracted special attention, as it exhibits intriguing properties \cite{Prins2014, Yichao2013, Jian2014, Andres2012, Jill2015, Yi2015}. Bulk  $\rm{MoS_2}$ is composed of covalently bonded S-Mo-S sheets that are bound by weak van der Waals forces and is an indirect band gap semiconductor with band gap of the order of 1.2 eV. However, a single layer of $\rm{MoS_2}$ is a direct band gap semiconductor with a band gap of 1.9 eV \cite{Rudren2014, Vikas2013}. This band gap lies in the visible range and the work function of $\rm{MoS_2}$ is compatible with the commonly used electrode materials.  Moreover, the single layer of $\rm{MoS_2}$ is fluorescent in nature with having quantum yield of the order of $10^4$ more than the bulk $\rm{MoS_2}$ \cite{Andrea2010}. Thus $\rm{MoS_2}$ has potential applications in electronic, optoelectronic and photonic devices \cite{Andrea2010, Kin2010, Rui2012}. For example, the field effect transistors based on $\rm{MoS_2}$ show high on/off switching ratio, \cite{Vikas2013} which is of the order of $10^8$. 

There are numerous studies on the intriguing properties and broad applications of $\rm{MoS_2}$ nanosheets; only recently, a study by Bruno et al. reported the  resonance Raman scattering in $\rm{MoS_2}$ nanosheets \cite{Carvalho2015}.    Conventionally, the EELS spectroscopy can be used to find out the band gap in semiconducting materials.  It has been studied  that the doping in semiconductors causes tuning in the band gap of these materials. In this work, we have studied the effect of different kinds of vacancies (Mo, S) in the EELS spectra of $\rm{MoS_2}$ nanosheets, and attempt has been made to correlated theoretical results of EELS with the Raman spectroscopic study. For the first time we have tried to correlate the theoretical results of EELS to the Raman spectroscopic results. It is found that molybdenum vacancy in  $\rm{MoS_2}$ nanosheets shows a EELS peak around 2.4 eV and Raman spectroscopic study shows enhancement of A$_{1g}$ peak at 532 nm laser excitation source, thus a resonance Raman scattering  phenomenon is  observed at 2.3 eV excitation source.    
	\section{Methods}
	\subsection{Computational methods}
	For the theoretical modeling and density functional theory (DFT) calculations, we have used the Vienna Ab Initio Simulation (VASP) code, \cite{kresse1993ab,kresse1994ab,kresse1996efficiency,kresse1996efficient} along with the MedeA software package. Before the evaluation of frequency dependent dielectric properties, all the structures were geometrically relaxed until the unbalanced   force     components converge below 0.02 eV/\AA. Simulations were performed under the generalized gradient approximation (GGA) with Perdew-Burke-Ernzerhof (PBE) exchange and correlation \cite{Perdew1996Generalized,Perdew1997Generalized}. A mesh cutoff energy of 400 eV, has been set in the expansion of plane wave basis sets and the electronic ground state convergence criteria has been set by $10^{-5}$ eV for all the systems. The Brillouin zone (BZ) has been sampled by $7 \times 7 \times 1$ Monkhorst-Pack (MP) grid point \cite{Monkhorst1976Special}.  Electron energy loss spectra (EELS) of pristine $\rm{MoS_2}$ system, $\rm{MoS_2}$ system with sulfur vacancy and $\rm{MoS_2}$ system with molybdenum vacancy have been explored in the framework of density functional theory (DFT).
	\subsection{Experimental methods}
	The proposed theoretical model is supported by Raman spectroscopic study of $\rm{MoS_2}$ sheets. The $\rm{MoS_2}$ nanosheets were prepared by chemical exfoliation method \cite{Yichao2015}. For this purpose, 1 g of bulk $\rm{MoS_2}$ powder was mixed with 1 mL of N-Methyl-2-pyrolidone (NMP) and the mixture was ground for 30 min in mortar pestle. The paste like mixture was put in vacuum oven at room temperature over night. The mixure was redispersed into 20 mL of NMP solvent and ultra-sonicated for 10 h using 25 W ultrasonication bath.  After ultrasonication, the mixture was centrifuged to separate out the  $\rm{MoS_2}$ nanosheets. The supernatant  was  collected from the solution which contained $\rm{MoS_2}$ nanosheets.  Such obtained dispersion of $\rm{MoS_2}$ sheets were drop cast on silicon substrate and annealed above 220 $^0$C for Raman and scanning electron microscopic (SEM) studies. For transmission electron microscopic (TEM) analysis the solution was put on carbon coated copper grid and dried at room temperature. 	
	
\section{Results and discussion}
\subsection{DFT calculation}

\begin{figure}
		\centering
		\includegraphics[width=0.6\linewidth]{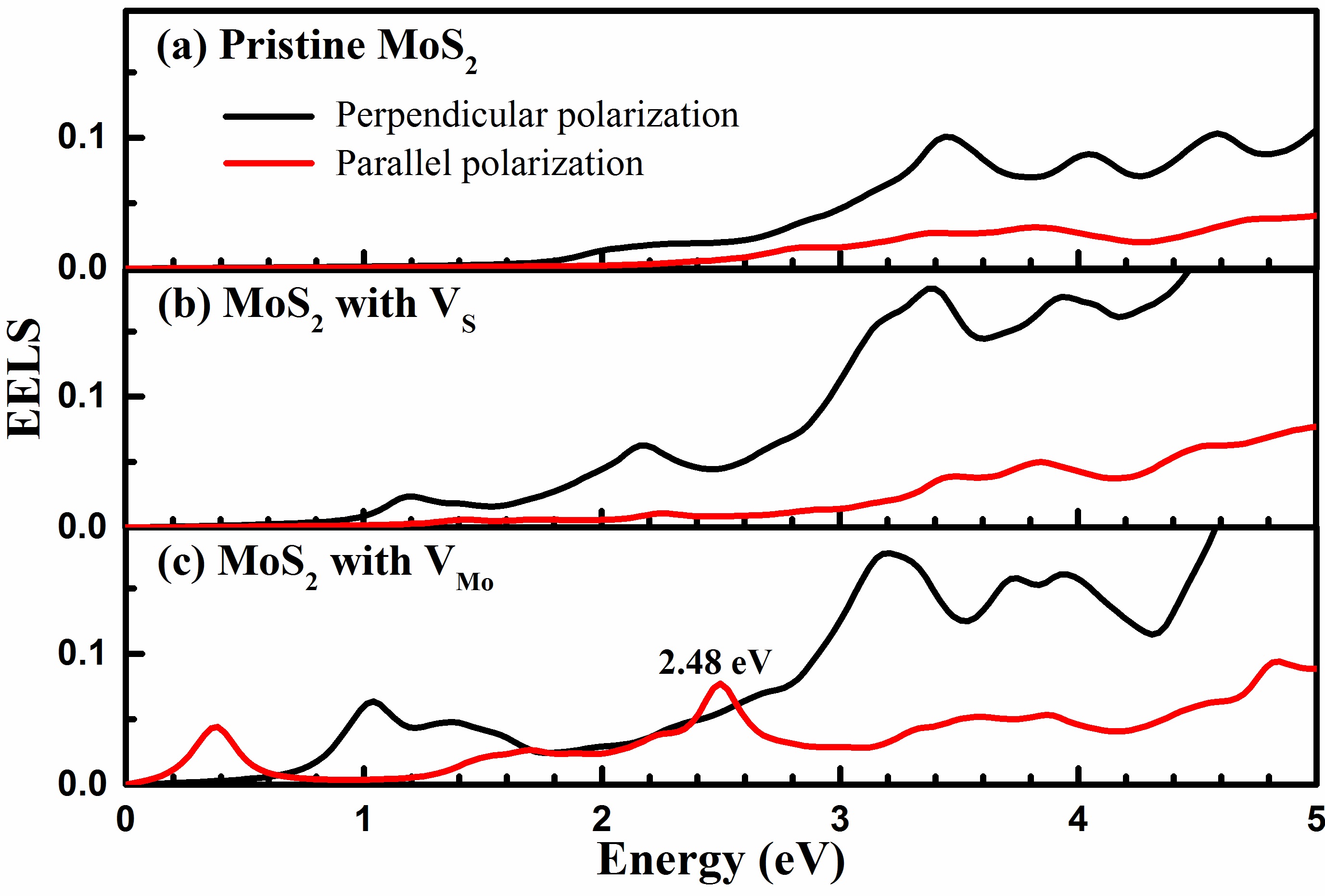} 
		\caption{Electron energy loss spectra of (a) pristine $\rm{MoS_2}$ nanosheets, (b) $\rm{MoS_2}$ nanosheets with sulfur vacancy ($\rm{V}_S$) and (c) $\rm{MoS_2}$ nanosheets with molybdenum vacancy ($\rm{V}_{Mo}$). The black line denotes perpendicular polarization and the red line denotes parallel polarization. \label{figure_EEL}}
	\end{figure}

In DFT calculations, first the frequency dependent dielectric function ($\epsilon(\omega)$) of a system has been computed; then electron energy loss specttra (EELS) of this system has been evaluated using the relation: $L(\omega)=Im(-1/\epsilon(\omega))$. From EELS (see figure \ref{figure_EEL}) it is identified that the spectrum shows asymmetrical nature in between two polarization states of electromagnetic (EM) wave; which is the manifestation of layered structure of $\rm{MoS_2}$. The results for the electronic energy loss spectra for $\rm{MoS_2}$ nanosheets are shown in  figure \ref{figure_EEL}. 

	Pristine MoS$_{2}$ system does not exhibit any significant loss peak in the visible range below 3.0 eV optical frequency. Besides, defect induced systems such as MoS$_{2}$ with sulfur vacancy as well as MoS$_{2}$ with molybdenum vacancy both demonstrate the existence of low energy loss peak in the visible part. Atomic vacancy induces localization of electron density near the vacancy site that actually gives rise to new plasmon excitations in the visible range as observed from EELS. One of the interesting features observed from EELS is that the emergence of new loss peaks in the infrared to visible part at 0.4 eV and 2.48 eV optical frequency for  perpendicular polarization (E$_{\bot}$) in case of molybdenum vacancy system (see figure \ref{figure_EEL} (c)). Besides, there exists no signature of such loss peaks in case of E$_{\bot}$ polarization in both pristine and sulfur vacancy system.
	
		\begin{figure} 
			\centering 
			\includegraphics[width=0.6\linewidth]{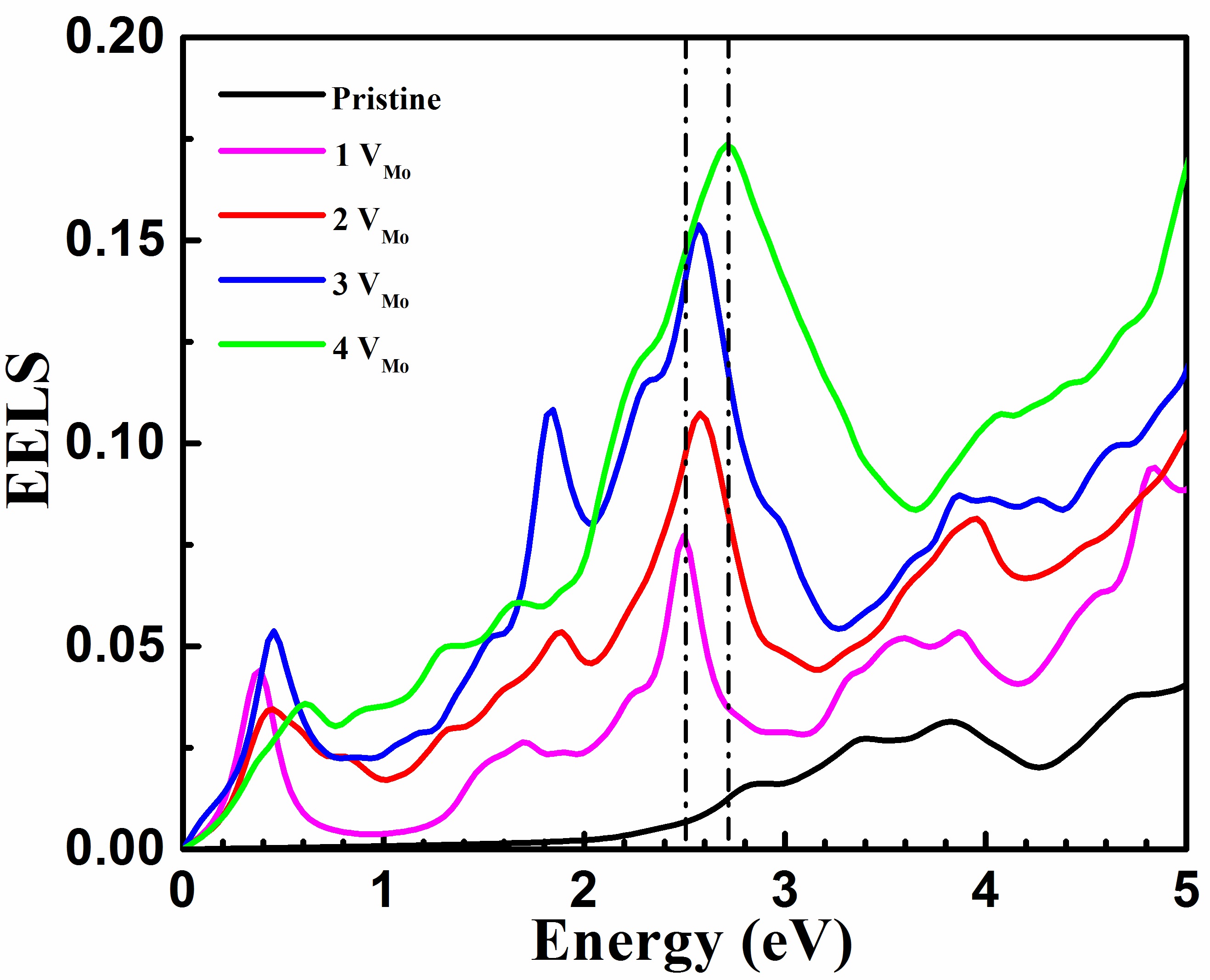} 
			\caption{Electron energy loss spectra of $\rm{MoS_2}$ nanosheets with varying concentrations of molybdenum vacancy ($\rm{V}_{Mo}$). The black line denotes parallel polarization for pristine $\rm{MoS_2}$. A single molybdenum vacancy (1$\rm{V}_{Mo}$) is depicted by the magenta line; 2$\rm{V}_{Mo}$ by red, 3$\rm{V}_{Mo}$ by blue, and 4$\rm{V}_{Mo}$ by green, respectively. \label{figure_EELS_variation}}
		\end{figure}
	
	Since $\rm{MoS_2}$ nanosheet is a two dimensional material with a layered structure, the in-plane (X-Y plane) symmetries are different from the out of plane (perpendicular direction) symmetries. Thus, the  associated chemical bonding and electron distribution is different along the perpendicular direction compared to the in-plane direction, which results in formation of two different excitation modes in the in-plane direction and perpendicular plane direction of the $\rm{MoS_2}$ nanosheet.   In general, they occur at different excitation energies also. The presence of $\rm{V}_{Mo}$ in the structure gives rise to electron localization at the vacancy site, on the $\rm{MoS_2}$ layer. These localized electrons result in a distinct, out-of-plane  excitation peak (or EELS peak) at $\sim$2.5 eV energy. Note that no excitation peak exists at the similar energy in case of in-plane plasma oscillation. To confirm the existence of the EELS peak at $\sim$2.5 eV energy, DFT calculations were performed for several systems having various molybdenum vacancies $\rm{V}_{Mo}$ or defect concentrations. Interestingly, it is noted that the increase of vacancy concentration enhances the EELS peak height (see figure \ref{figure_EELS_variation}). Moreover, a little blue shift of the peak position is observed with increasing  defect concentration, although the shift does not affect the final results and conclusions of this present observation.

		
	\subsection{TEM analysis}
\begin{figure}
		\centering
		\includegraphics[scale=0.3]{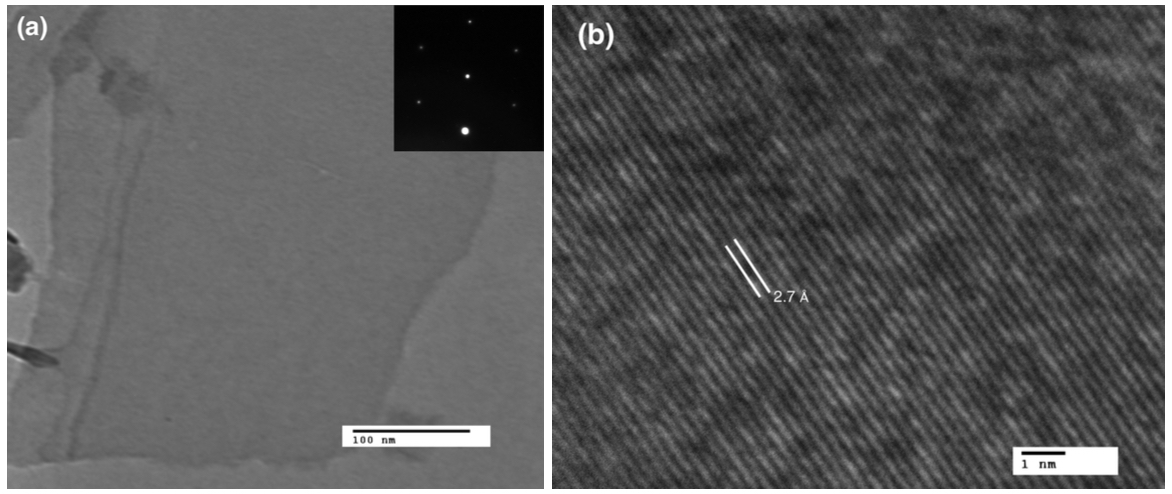} 
		\caption{TEM images of (a)  $\rm{MoS_2}$ nanosheets, with inset showing SAED pattern (scale equals to 100 nm); (b) lattice spacing in $\rm{MoS_2}$ nanosheet (scale equals 1 nm). \label{figure_TEM}}
	\end{figure}

	Presence of sheets like structure in $\rm{MoS_2}$ was confirmed by  transmission electron microscope  image analysis. The TEM images were obtained by a transmission electron microscope (model JEOL-2100F, Japan), operating at 200 kV. The TEM image of $\rm{MoS_2}$ nanosheets is shown in figure \ref{figure_TEM}. 
	
	The inset of figure \ref{figure_TEM}(a) shows the selective area electron diffraction (SAED) pattern from $\rm{MoS_2}$ nanosheet. Figure \ref{figure_TEM}(b) shows the lattice spacing in $\rm{MoS_2}$ nanosheet. The lattice spacing between adjacent planes   is   of the order of 2.7 {\AA}  which corresponds to (100) plane.	
	\subsection{SEM analysis}
	\begin{figure} 
		\centering
		\includegraphics[width=0.5\linewidth]{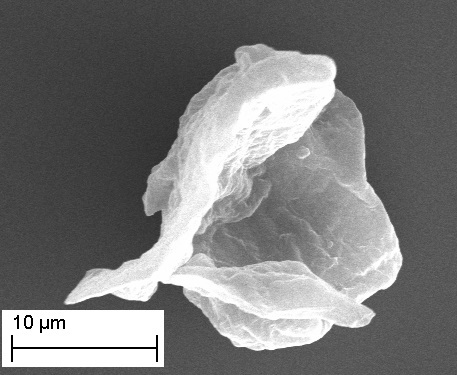} 
		\caption{SEM image of $\rm{MoS_2}$ nanosheets.   \label{figure_SEM}}
	\end{figure}
	The SEM image of MoS$_2$ nanosheets was recorded by scanning electron microscope (Zeiss EVO40). 
	The sheets like structure of MoS$_2$ was observed in the scanning electron microscopy. The SEM image of MoS$_2$ nanosheets is shown in figure \ref{figure_SEM}. The size of these nanosheets are of order of 10 $\mu$m.

\subsection{Raman spectroscopic study}	
	The Raman spectra of $\rm{MoS_2}$ were collected by Renishaw inVia  confocal Raman spectrometer using four different excitation laser sources 405 nm, 532 nm, 632 nm and 785 nm. For 405 nm and 532 nm lasers the associated grating is 2400 lines per mm while for 632 nm and 785 nm lasers it is 1200 lines per mm. The Raman spectra were collected by using 50X objective lens having numerical aperture 0.25.   
	
		\begin{figure}
			\centering
			\includegraphics[width=0.8\linewidth]{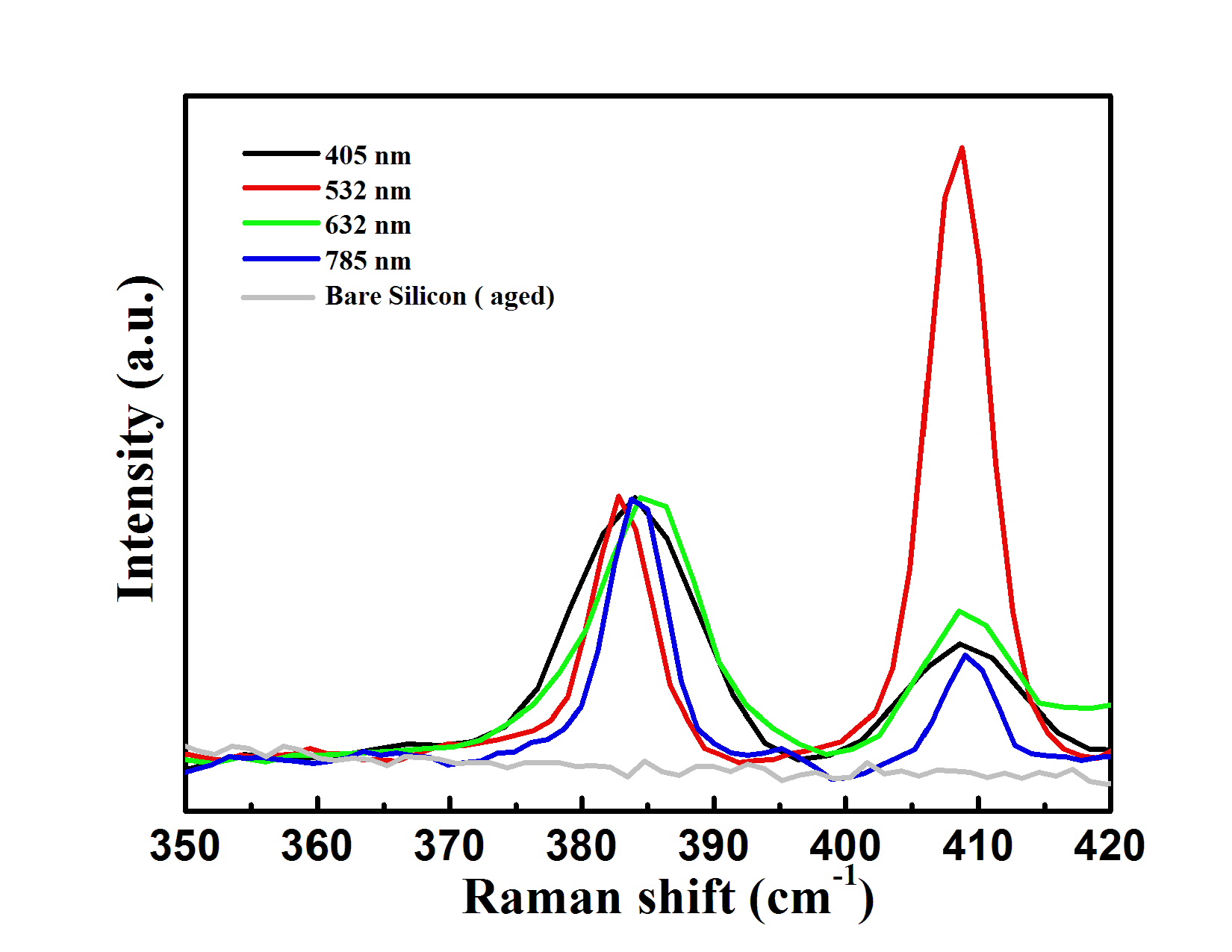} 
			\caption{Raman spectra of $\rm{MoS_2}$ nanosheets for different laser excitation sources: 405 nm  (3.06 eV), 532 nm (2.33 eV), 632 nm (1.96 eV) and 785 nm (1.58 eV). The Raman spectrum of bare silicon (aged) is plotted in grey.  \label{figure_raman}}
		\end{figure}

	Figure \ref{figure_raman} shows the Raman spectra of $\rm{MoS_2}$ nanosheets. The spectra consist two Raman peaks one at 385 cm$^{-1}$ and other at 408 cm$^{-1}$. The Raman peak at 385 cm$^{-1}$ is known as $E_{2g}^{1}$ peak and other at 408 cm$^{-1}$ is $A_{1g}$ peak, these peaks originate due to in plane and out of plane vibrations of S-Mo-S bonds. The effect of excitation source on the Raman peaks of $\rm{MoS_2}$ nanosheets has been analyzed. For all three wavelengths of laser sources, 405 nm (3.06 eV), 632 nm (1.96 eV) and 785 nm (1.58 eV)  the intensity of $E_{2g}^{1}$ peak is greater than the intensity of $A_{1g}$ peak while in case of 532 nm (2.33 eV) excitation source the intensity of $A_{1g}$ is more than that of the $E_{2g}^{1}$ peak. Our theoretical analysis shows that an EELS  peak is observed around 2.48 eV. Hence the enhancement of the Raman peak is due to the resonance occurring at 532 nm.
	
	In order to verify whether  the intensity enhancement of $A_{1g}$ Raman peak under 532 nm laser excitation,  comes from the Si/$\rm{SiO_2}$ substrate, we tested the Raman response in the bare area of the sample (without the $\rm{MoS_2}$) to see Si/$\rm{SiO_2}$ substrate. Here, we have used aged Si, since it may have more native $\rm{SiO_2}$. As evident from figure \ref{figure_raman}, there is no conflict of $\rm{SiO_2}$ Raman peaks within the region of $\rm{MoS_2}$ peaks and the intensity enhancement of the $A_{1g}$ peak (identified at 2.48 eV in the visible range for $\rm{MoS_2}$ nanosheet with molybdenum vacancy) comes from a resonance Raman scattering, and certainly not from the Si/$\rm{SiO_2}$ substrate.
		\begin{figure} 
			\centering 
			\includegraphics[width=0.6\linewidth]{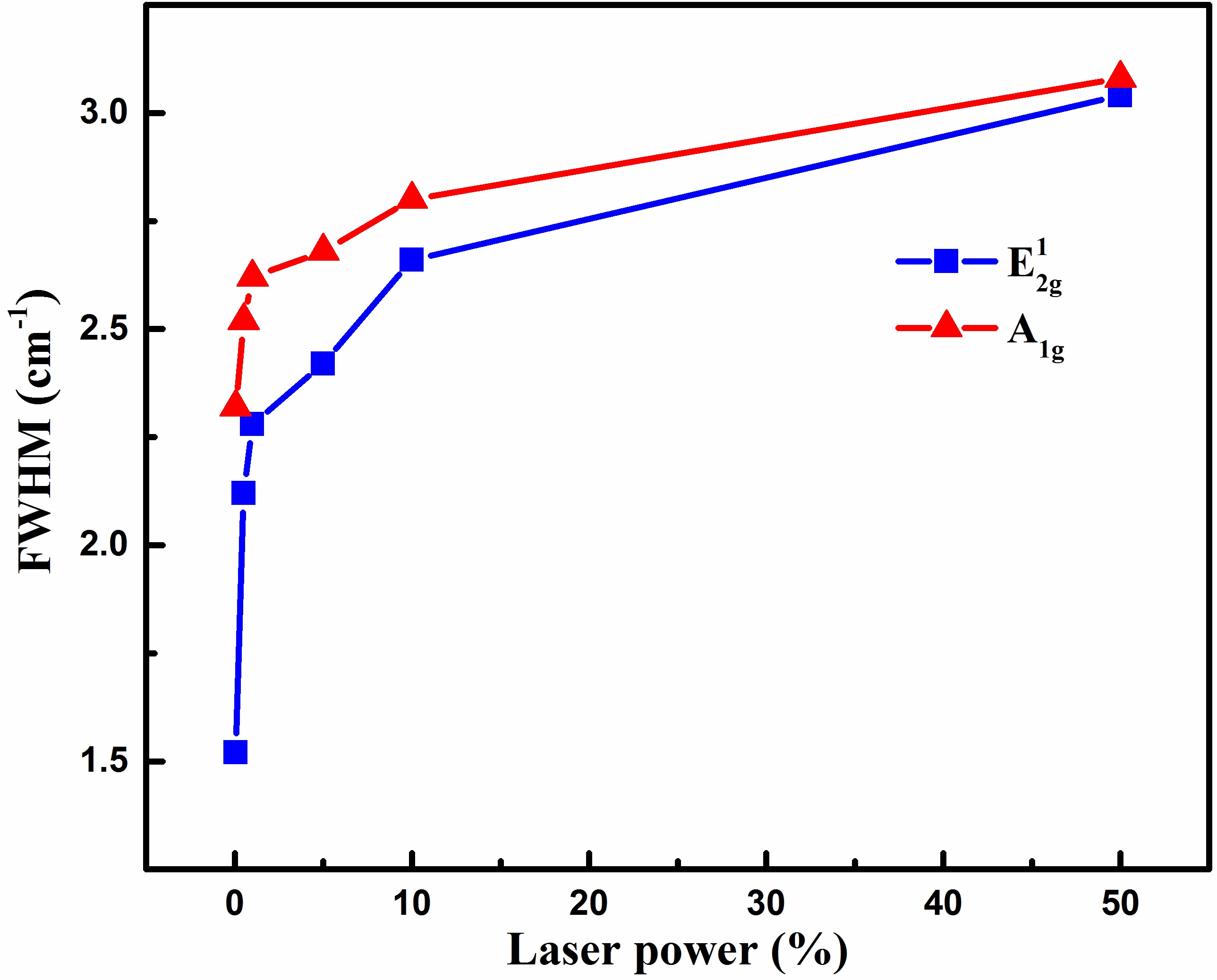} 
			\caption{Effect of laser (532 nm) power on the full width half maximum (FWHM) of $E_{2g}^{1}$ and $A_{1g}$ peaks of $\rm{MoS_2}$ nanosheets.   \label{figure_FWHM}}
		\end{figure}
	
	The effect of laser power on the full width half maximum (FWHM) of both the peaks have been studied using 532 nm laser source (figure \ref{figure_FWHM}). With increase in the power of laser the FWHM value of $E_{2g}^{1}$ and $A_{1g}$ peaks increases and attains a saturation level.

	\section{Conclusions}
	In summary, we have demonstrated the presence of EELS peak in molybdenum disulfide ($\rm{MoS_2}$) nanosheets in the visible energy range. Theoretically, using DFT simulations and calculations, an EELS peak was identified at 2.48 eV (visible range), for $\rm{MoS_2}$ nanosheets with molybdenum vacancy; for pristine  $\rm{MoS_2}$ and $\rm{MoS_2}$ with sulfur vacancy, the EELS did not show any resonance peaks for parallel polarization.
	This was then confirmed experimentally, using four different laser sources for the Raman scattering study of $\rm{MoS_2}$ nanosheets. In the cases of three laser sources with wavelengths 405 nm (3.06 eV),  632 nm (1.96 eV) and  785 nm (1.58 eV), respectively, the intensity of $E_{2g}^{1}$  Raman peak was more than the $A_{1g}$ Raman peak, while in the case of excitation source of 532 nm (2.33 eV), the intensity profile was reversed and $A_{1g}$ peak was the most intense. Such a resonance phenomena may trigger some new features such as generation of localized energy states. 	The presented 2D nanosheets have a great potential for future  optical systems, especially in nanophotonic devices operating at visible wavelengths.\\
	\\
	$\bf{Acknowledgement}$\\ \\	
	AC acknowledges the financial support by institutional research funding IUT (IUT39-1) of the Estonian Ministry of Education and Research. AC and ASP acknowledge financial support from grant number BT/BI/03/004/2003(C) of Government of India, Ministry of Science and Technology, Department of Biotechnology, Bioinformatics division.
	Authors thank the AIRF, JNU for TEM and SEM characterizations and the FIST (DST, Govt. of India) UFO scheme of IIT Delhi for Raman/PL facility.

\section*{References}
\bibliography{main}
\bibliographystyle{iopart-num}

\end{document}